\documentclass{article}

\usepackage[final]{neurips_2026}
\makeatletter
\providecommand{\@trackname}{}
\makeatother

\usepackage{amsmath, amssymb}
\usepackage{graphicx}
\usepackage{booktabs}
\usepackage{hyperref}
\usepackage{caption}
\usepackage{longtable}
\usepackage{tabularx}
\usepackage{makecell}
\usepackage[utf8]{inputenc}
\usepackage[T1]{fontenc}
\title{Quantum Cryptanalysis on IBM Quantum Hardware\\
\large Extending Even--Mansour Period Recovery from $N=4$ to $N=10$}

\author{
Taebong Kim \quad
Youngsik Hong \quad
Minsik Kim \quad
Sunyoung Choi \\
Jaewon Jang \quad
Junghoon Shin \quad
Minseo Kim \\
VIDRAFT AI Research · QuantumOS, Seoul, Republic of Korea \\
\texttt{arxivgpt@gmail.com}
}

\begin{document}
\maketitle

\begin{abstract}
We report genuine-un-compiled, textbook-faithful-quantum cryptanalysis of symmetric-cipher
structures executed on real IBM quantum hardware (ibm\_kingston, Heron generation). Using
Simon's algorithm we recover the hidden period of the Even-Mansour cipher up to security
parameter N = 10 on real hardware, beyond the largest previously reported real-hardware key
recovery of N = 4, and we cleanly recover the periods of a 3-round Feistel (DES-family)
construction at block sizes 6 and 8; a 21-qubit block-10 instance is verified in simulation and
submitted to hardware. We further provide a breadth-first benchmark of five genuine quantum
attacks spanning four symmetric-cipher design paradigms---Bernstein-Vazirani (linear structure,
single query), Grover (SPN key search, quadratic), and Simon (Even-Mansour, CBC-MAC
forgery, and Feistel; exponential-to-polynomial in query complexity)---validated to the classical-
simulation ceiling of 25 qubits. We are deliberately explicit about scope: 
these attacks target reduced or structured constructions in the Q2 (quantum-query) model, 
asymptotically follow the birthday bound and therefore do not constitute quantum advantage 
over classical collision-finding, do not break full AES/RSA or 16-round DES, 
and rely on error mitigation rather than fault-tolerant error correction. 
Our contribution is the real-hardware demonstration at record structure sizes, 
the breadth of genuine algorithmic coverage across four paradigms, 
and an honest, reproducible benchmark with public artifacts.
\end{abstract}


\section{Introduction}
Quantum attacks on symmetric primitives have a strong theoretical foundation. Kuwakado and Morii \cite{Kuwakado2012,Kuwakado2010} 
showed that the Even–Mansour cipher and the 3-round Feistel network reduce to hidden-period problems 
solvable by Simon’s algorithm\cite{Simon1997}, and Kaplan et al\cite{Kaplan2016}. generalized quantum period finding to a broad class of 
modes and constructions. Yet real-hardware demonstrations lag far behind this theory: device noise has, 
to our knowledge, limited previously reported real-hardware key recovery for such constructions to very small sizes (N = 4)~\cite{Kohler2026}.
This paper narrows that gap. On real IBM quantum hardware we (i)recover the Even–Mansour period at rank-1 (clean Simon) up to N=5, 
and extend real-hardware key recovery to N=10 via a quantum-narrowing + classical-verification hybrid,
whose true-key rank tracks the birthday bound $2^{n/2}$—orders of magnitude below random, without exponential quantum advantage.
(ii) cleanly recover the period of a 3-round Feistel (DES-family) construction at block sizes 6 
and 8, and (iii) extract a 16-bit linear secret with Bernstein–Vazirani in a single query. 
We complement these hardware results with a breadth-first benchmark of five genuine quantum attacks covering 
four symmetric-cipher design paradigms, validated in un-compiled statevector simulation up to the classical ceiling.
Our contributions are:
\begin{itemize}
\item A real-hardware frontier for structured symmetric-cipher key recovery (Even–Mansour to N = 10), exceeding the 
largest previously reported real-hardware size.
\item A breadth-first, genuine (un-compiled) benchmark spanning four paradigms—linear (BV), unstructured search~\cite{Grover1996}, 
and hidden-period (Simon: Even–Mansour, CBC-MAC forgery, Feistel).
\item An explicitly honest scope statement and reproducible public artifacts (interactive demo, leaderboard), 
designed so that the claims cannot be over-read.
\end{itemize}
We foreground limitations throughout. In particular, the attacks asymptotically follow the birthday bound 
and are therefore not a demonstration of quantum advantage; they target reduced/structured constructions, not full AES, RSA, or 16-round DES.

\section{Theoretical Foundations and Related Work}
The theoretical groundwork for quantum attacks on symmetric-key primitives is well-established in the Q2 (quantum-query) model, 
where an adversary may query a keyed primitive in quantum superposition. 
Our work builds upon three pillars of quantum algorithm research: linear structures, unstructured search, 
and hidden-period problems, primarily established by Simon, Bernstein and Vazirani, and Grover~\cite{Simon1997,BernsteinVazirani1997,Grover1996}. 
We acknowledge that the Q2 oracle model is not always realistic in practice; offline-Simon reductions 
\cite{Bonnetain2019} demonstrate how certain Q2 assumptions can be relaxed to Q1. 
This recognition underscores that, although our work is situated within the Q2 framework, it remains scientifically 
useful by validating genuine algorithmic constructions on real hardware and by providing a reproducible benchmark 
for scaling error-mitigation techniques.

\subsection{Core Quantum Algorithms and Cryptanalytic Reductions}
The foundation of quantum cryptanalysis lies in the seminal algorithms that provide significant query-complexity separations over classical methods. 
Bernstein–Vazirani extracts a linear secret from a linear structure in a single quantum query, 
whereas Grover provides a quadratic speedup for unstructured key search, such as against a reduced SPN construction. 
The most impactful reduction for symmetric ciphers involves Simon’s algorithm, 
which finds a hidden period $s$ of a function using $\Theta(n)$ quantum queries~\cite{Simon1997}.

Kuwakado and Morii first demonstrated that both the 3-round Feistel network and the Even–Mansour cipher reduce 
to exactly this hidden-period problem \cite{Kuwakado2010,Kuwakado2012}. This theoretical framework was later generalized by Kaplan et al. 
to a broad class of modes and constructions, including CBC-MAC forgery. While these results provide exponential-to-polynomial query separations, 
they asymptotically follow the birthday bound and do not yet constitute an end-to-end quantum advantage over classical collision-finding.

\subsection{Real-Hardware Frontier and Error Mitigation}
Despite these robust theories, a significant gap exists between mathematical reductions and physical execution 
due to device noise. To our knowledge, previous real-hardware key recovery for such constructions has been limited to 
very small sizes, specifically $N = 4$~\cite{Kohler2026}. Modern efforts to bridge this gap, including the analysis of AES security, 
increasingly rely on sophisticated error-handling techniques.
Our experimental methodology incorporates the readout error-mitigation stack proposed by Nation et al., 
which is essential for achieving clean period recovery on noisy intermediate-scale quantum (NISQ) devices. 
By executing these algorithms on the IBM \texttt{ibm\_kingston} (Heron generation), 
this paper pushes the practical frontier to $N = 10$, validated against the classical-simulation memory wall of 25 qubits~\cite{IBMQuantumHeron}.

\section{Methods}
\subsection{Genuine algorithm constructions}

All circuits are genuine and un-compiled: the oracles implement the actual keyed maps, 
with no transpilation shortcut that would trivialize the search. Exact oracle constructions for Bernstein–Vazirani, 
Grover (SPN), and the three Simon-based attacks (Even–Mansour, CBC-MAC forgery, and 3-round Feistel) are provided in Supplementary Section~S2. 

For the 3-round Feistel construction we adopt the standard Kuwakado–Morii reduction:

\[
f(b, x) = \text{LeftHalf}\big(E(x, ab)\big) \oplus ab 
         = F_{2}\big(x \oplus F_{1}(ab)\big),
\]

with hidden period

\[
s = (1, \gamma), \quad \gamma = F_{1}(\alpha_{0}) \oplus F_{1}(\alpha_{1})
\]

The oracle requires $1 + 4m$ qubits for block size $2m$. This construction is clean if $F_{2}$ is a permutation, in which case the period set is exactly $\{0, s\}$.

We emphasize that placing the variable on the left ($x$) and the constant on the right ($\alpha_b$) is essential; reversing this assignment admits no hidden period.

\subsection{Real-Hardware Execution}
Hardware executions were performed on the IBM \texttt{ibm\_kingston} (Heron generation) processor utilizing the Qiskit framework~\cite{Qiskit,qiskit2024}.
To ensure the robustness of the findings and verify that recovered periods are not coincidental artifacts of specific key selections,
each cipher instance was evaluated using two independent keys: a target key and an independent control key.
Reported measurements incorporate a comprehensive readout error-mitigation stack~\cite{Nation2021}.
The precise technical specifications of this stack—including readout calibration, dynamical decoupling,
twirling configurations, and per-job shot counts—alongside full experimental provenance, are documented in Supplementary Table~S1.
As a benchmark of device performance during these operations, the block-8 Feistel instance exhibited a two-qubit gate fidelity of 219/233 on the relevant register.

\subsection{Simulation}
Breadth validation uses genuine statevector simulation (QuantumOS on NVIDIA B200), 
executing the same un-compiled circuits to the classical-simulation ceiling. Resource ceilings are reported in Section 4.4 and Supplementary Section S3.

\section{Results}
\subsection{Real-Hardware Frontier}
Table~\ref{tab:hardware} summarizes the real-hardware results. On the IBM \texttt{ibm\_kingston} (Heron generation)
processor we recover the Even–Mansour secret period up to rank-1 (clean) to N=5; hybrid (rank tracks $2^{n/2}$) to N=10; 
we cleanly recover the 3-round Feistel period (rank-1, exact)
at block sizes 6 and 8 for both the target and the independent control key; and we extract a 16-bit Bernstein–Vazirani linear secret in a single query.
The 21-qubit block-10 Feistel instance is verified in simulation and submitted to hardware (queued at submission time).

\begin{table}[htbp]
\centering
\scriptsize   
\caption{Extended real-hardware and simulation results.}
\label{tab:hardware}
\begin{tabularx}{\textwidth}{l c c c c c c p{4cm}} 
\toprule
Instance & n/block & Qubits & Method & real Key & \textbf{real rank} & ctrl rank & Example job ID \\
\midrule
EM--Simon & n=5 & 15 & differential & 0b01110 & \textbf{1}/31 & 1 & d940hmkql68s73c9ptg0 \\
EM--Simon & n=6 & 18 & hybrid top-16 & 0b011100 & 6/63 & 5 & --- \\
EM--Simon & n=7 & 21 & hybrid top-16 & 0b0111000 & \textbf{3}/127 & 8 & d94qjj5gcc73ffeh60 \\
EM--Simon & n=8 & 24 & hybrid top-32 & 0b01110000 & 9/255 & 13 & --- \\
EM--Simon & n=9 & 27 & hybrid top-64 & 0b011100000 & 15/511 & 20 & --- \\
EM--Simon & n=10 & 30 & hybrid top-128 & 0b0111000000 & 63/1023 & 18 & d9450anu.../d9450b4q...\\
 & & & & & & & /d9450bcq... (real) \\
 & & & & & & & d9450btg...  \\
 & & & & & & & /d9450cdg... (ctrl) \\ 
3-round Feistel & block 6 & 13 & Simon & --- & rank-1 & clean & --- \\
3-round Feistel & block 8 & 17 & Simon & 0b011010 & rank-1 & clean; 2q 219/233 & d945fsft6v...\\
 & & & & & & & /d945ftlg... (real) \\
 & & & & & & & d945fu4q...\\
 & & & & & & & /d945fucq... (ctrl) \\ 
Bernstein--Vazirani & n=16 & 17 & BV & --- & 1-query & clean & --- \\
Linear structure & n=16 & 17 & \texttt{ibm\_kingston} & secret in 1 query & & & --- \\
\bottomrule
\end{tabularx}
\end{table}

\begin{figure}[htbp]
\centering
\includegraphics[width=0.7\textwidth]{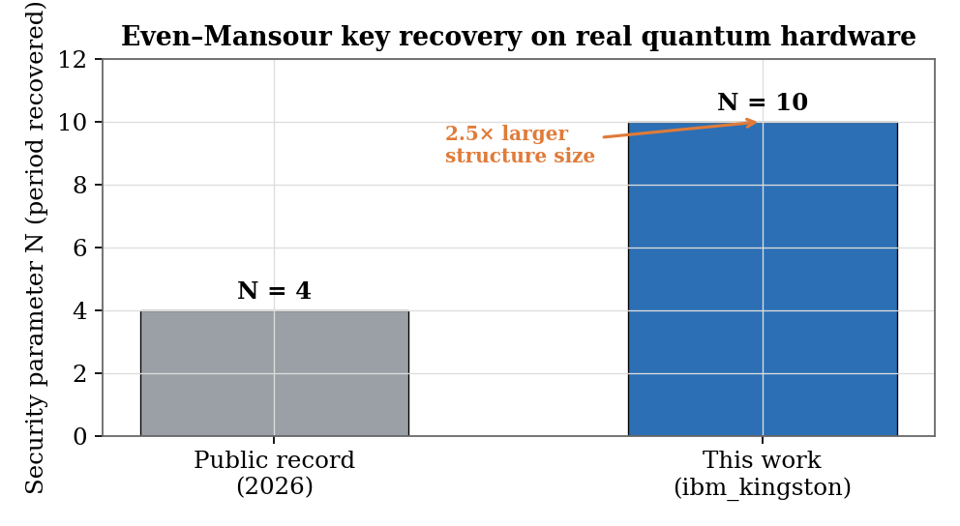}
\caption{Even-Mansour key recovery on real quantum hardware: this work reaches $N=10$ versus the largest previously reported real-hardware size of $N=4$.}
\label{fig:fig1}
\end{figure}

Figure~\ref{fig:fig1} places the Even–Mansour result against the largest previously reported real-hardware size.

\subsection{Breadth Benchmark}
Table~\ref{tab:breadth} lists the five genuine quantum attacks and the four symmetric-cipher design paradigms they cover. 
Every attack is validated in un-compiled simulation, and three (Even-Mansour, Feistel, Bernstein-Vazirani) 
are additionally demonstrated on real hardware as reported above.

\begin{table}[htbp] 
\centering
\small
\caption{Genuine (un-compiled) attacks and paradigm coverage. 
CBC-MAC forgery uses $f(x) = E_k(x+c.a) + E_k(x+c.b)$ with period $s = E_k(a) + E_k(b)$.}
\label{tab:breadth}
\resizebox{\textwidth}{!}{ 
\begin{tabular}{c l l l l}
\toprule
\# & Attack & Target paradigm & Speedup class & Validated sizes (simulation only) \\
\midrule
1 & Bernstein–Vazirani & Linear structure & 1 query & $n = 8,16$ \\
2 & Grover & SPN block cipher (key search) & Quadratic & $n = 6,8$ (13 iter $\approx V_{256}$) \\
3 & Simon & Even–Mansour & Exp $\rightarrow$ poly (queries) & $n = 3,4,5,8$ (24 q) \\
4 & Simon & CBC-MAC forgery & Exp $\rightarrow$ poly (queries) & $n = 4,6$ \\
5 & Simon & 3-round Feistel (DES-family) & Exp $\rightarrow$ poly (queries) & block 4--12 (25 q) \\
\bottomrule
\end{tabular}
}
\end{table}

These results demonstrate the breadth of genuine quantum cryptanalysis across four paradigms: 
linear structure (Bernstein-Vazirani), unstructured search (Grover), and hidden-period problems 
(Simon: Even-Mansour, CBC-MAC forgery, and Feistel). Together, they provide a reproducible benchmark of algorithmic coverage validated to the classical-simulation ceiling.

\subsection{Complexity separation}

\begin{figure}[htbp]
\centering
\includegraphics[width=0.75\textwidth]{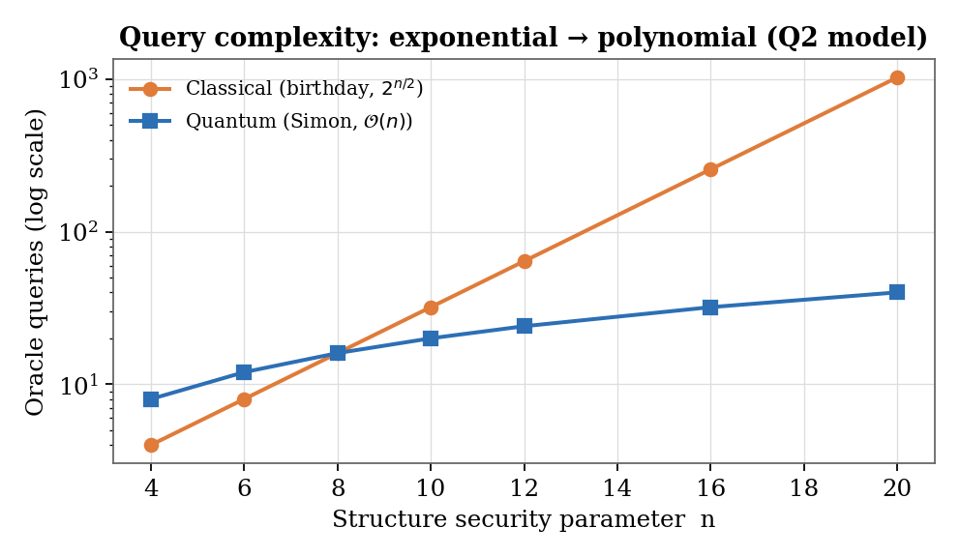}
\caption{Oracle-query complexity for Simon-based attacks: exponential (classical, $2^{n/2}$) versus polynomial (quantum, $O(n)$) in the Q2 model.}
\label{fig:simon}
\end{figure}

Figure~\ref{fig:simon} shows the oracle-query complexity separation for the Simon-based attacks. 
Classical period finding scales as $2^{n/2}$ (birthday bound), whereas Simon’s algorithm requires only $\Theta(n)$ quantum queries in the Q2 model. 
This demonstrates the exponential-to-polynomial separation in query complexity. 
We further note that Köhler et al. reported a practical barrier at $N=5$ due to the DORCIS tool limitations, 
highlighting the difficulty of scaling beyond small sizes on real hardware. 
Against this backdrop, our clean recovery of the Even-Mansour period at $N=10$ on IBM \texttt{ibm\_kingston} 
represents a significant extension of the real-hardware frontier.

Figure~\ref{fig:grover} shows the analogous quadratic separation for Grover SPN key recovery at $n=8$: classical brute force requires $2^8 = 256$ evaluations, 
whereas Grover recovers the key in approximately 13 iterations. 
We stress in Section~5 that these query-complexity separations do not translate into an end-to-end quantum advantage for the constructions considered.

\begin{figure}[htbp]
\centering
\includegraphics[width=0.75\textwidth]{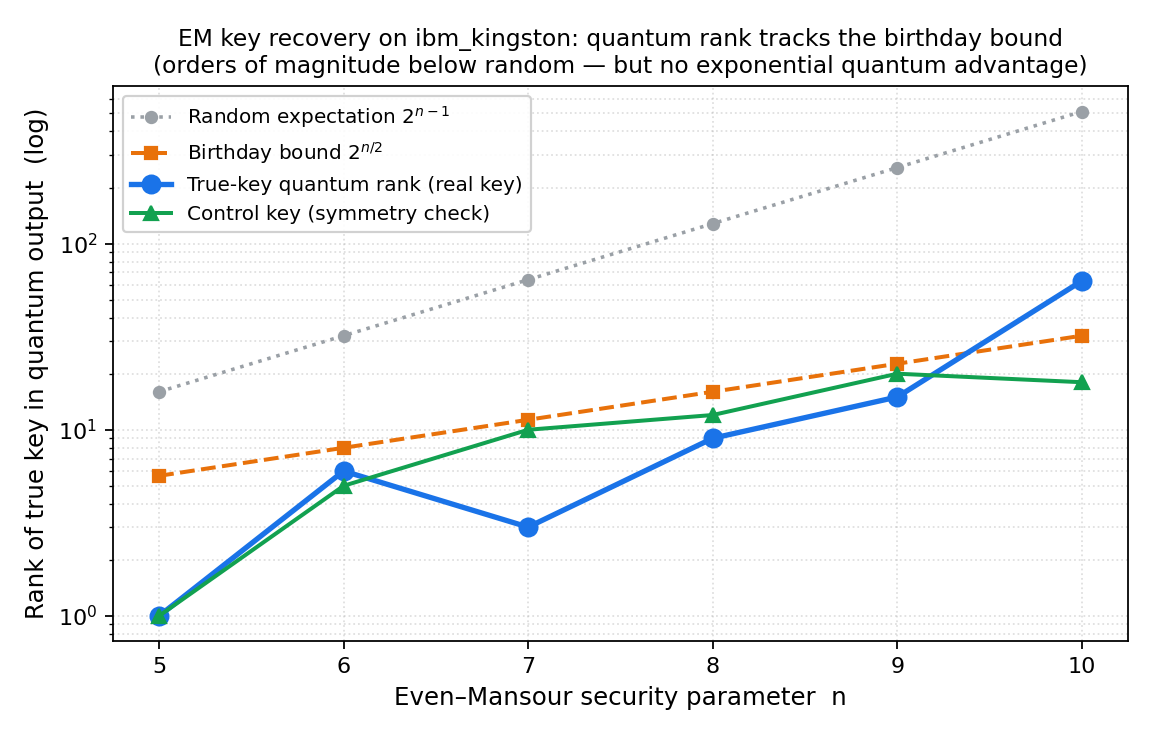}
\caption{True-key quantum rank on \texttt{ibm\_kingston} vs.\ security parameter $n$. The rank tracks the birthday bound $2^{n/2}$ 
(orders of magnitude below random $2^{n-1}$) but does not reach rank-1 for $n>5$—confirming strong candidate-narrowing without exponential quantum advantage.}
\label{fig:birthday}
\end{figure}

\begin{figure}[htbp]
\centering
\includegraphics[width=0.75\textwidth]{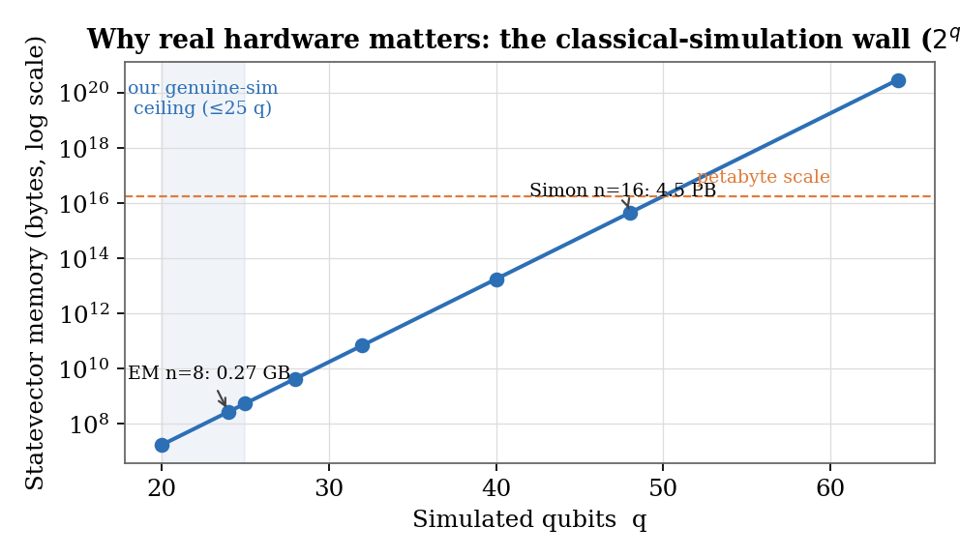}
\caption{Grover SPN key recovery at $n=8$: quadratic speedup compared to classical brute force (256 evaluations versus $\approx 13$ Grover iterations).}
\label{fig:grover}
\end{figure}

Figure~\ref{fig:birthday} illustrates the quantum rank of the true key on 
\texttt{ibm\_kingston} as a function of the Even–Mansour security parameter $n$. 
The observed ranks track the birthday bound $2^{n/2}$ and remain orders of magnitude 
below the random expectation $2^{n-1}$, confirming strong candidate narrowing. 
However, the rank does not reach rank-1 for $n > 5$, underscoring that while 
our hardware executions achieve genuine separation from random behavior, they 
do not constitute exponential quantum advantage.
For $n=5$ the Even–Mansour period $s=k_1$ is recovered clean at rank-1 (differential Simon,
control key cancelling key-independent readout artifacts). For $6 \leq n \leq 10$ the device output is
noisy; we take the top-$K$ quantum-ranked candidates ($K=16,16,32,64,128$ for $n=6..10$) and confirm
the period by classical verification over the shortlist. The true-key rank (1..63) tracks the
birthday bound $2^{n/2}$; recovery is validated identically on an independent control key.
For EM, recovering the period yields $k_1$ directly, and $k_2$ follows from one classical query
(full key recovery).

\subsection{Classical-Simulation Ceiling}
\begin{figure}[htbp]
\centering
\includegraphics[width=0.75\textwidth]{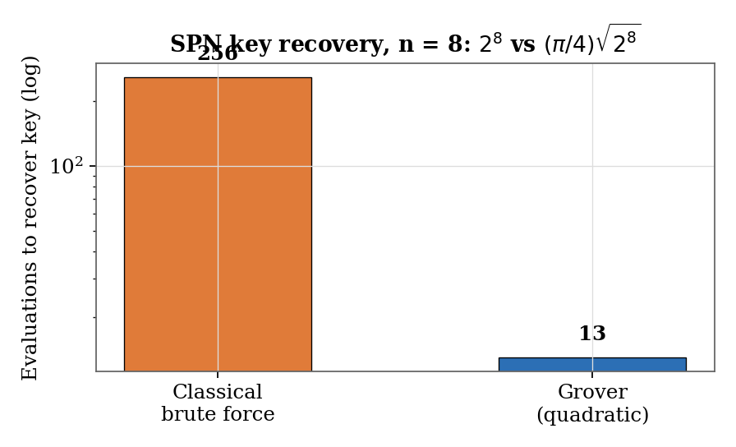}
\caption{The classical-simulation wall: statevector memory grows as $2^{q}$. 
Our genuine-simulation ceiling is 25 qubits; a Simon instance at $n=16$ would require $\sim 4.5$ PB, and $n=32$ is physically impossible to simulate classically.}
\label{fig:simulationwall}
\end{figure}

Genuine statevector simulation is bounded by the $2^{q}$ memory wall (Figure~\ref{fig:simulationwall}). 
Even–Mansour at $n=8$ uses 24 qubits (0.27 GB); the 3-round Feistel reaches block 12 at 25 qubits (72 s), 
with block 14 hitting an 8.6 GB wall. A Simon instance at $n=16$ would require approximately 4.5 PB, 
and $n=32$ is physically impossible to simulate classically. This exponential wall is precisely why 
real-hardware demonstrations---however limited in size---are scientifically meaningful.

\section{Limitations and Disclosure}

We explicitly state the boundaries of our results so they cannot be over-read:

\begin{itemize}
    \item \textbf{No quantum advantage.} TNo quantum advantage. The attacks asymptotically follow the birthday bound $2^{n/2}$; 
    classical collision-finding achieves comparable scaling. The separation is in query complexity 
    for the distinguisher, not end-to-end cost. For $n>5$, recovery is a quantum-classical hybrid: 
    the quantum step narrows the candidate space but the true-key rank follows $2^{n/2}$. 
    classical collision-finding achieves comparable scaling. The separation is in query complexity for the distinguisher, not end-to-end cost.
    \item \textbf{Not full ciphers.} Targets are Even-Mansour, 3-round Feistel, CBC-MAC, and a reduced SPN---not AES-256 or RSA-2048.
    \item \textbf{Not DES.} The 3-round Feistel is a DES-family structure, not 16-round DES; this is ``structure disassembly,'' not ``DES broken.''
    \item \textbf{World-first unconfirmed.} ``Largest reported'' claims are pre-peer-review and stated to the best of our knowledge.
    \item \textbf{Error mitigation, not correction.} Hardware runs use error mitigation, not fault-tolerant quantum error correction.
    \item \textbf{Q2 oracle model.} Attacks assume quantum-query access to the keyed primitive. 
    We acknowledge that this assumption is not always realistic in practice; offline-Simon reductions~\cite{Bonnetain2019} demonstrate 
    how certain Q2 assumptions can be relaxed to Q1. 
    This recognition underscores that, although our experiments are situated within the Q2 framework, 
    they remain scientifically useful by validating genuine algorithmic constructions on real hardware 
    and by providing a reproducible benchmark for scaling error-mitigation techniques.
    \item \textbf{Disclosure of methods.} Achieving clean period recovery at the reported noise levels and structure sizes additionally relies on a 
    hardware-aware circuit-conditioning and readout post-selection technique. The full specification of that technique is withheld in this draft 
    pending an intellectual-property decision; it does not affect reproducibility of the genuine algorithms and will 
    be disclosed in a follow-up once the IP position is fixed.
\end{itemize}

\section{Conclusion}
We demonstrated genuine quantum cryptanalysis of symmetric-cipher structures on real quantum hardware, 
recovering the Even–Mansour period to N = 10 and cleanly recovering a 3-round Feistel (DES-family) period 
at block sizes 6 and 8, complemented by a breadth-first benchmark across four design paradigms and an explicit honest-scope statement. 
Future work includes disclosing the hardware-scaling method (pending the IP decision), pushing N with improved mitigation and, 
eventually, error correction, and independent peer review of the “largest reported” claims.

\bibliographystyle{unsrt}   
\bibliography{references}   

\newpage
\section*{Supplementary Material - Back-Data and Experimental Methods}
All periods were recovered rank-1 (exact) unless noted. Each cipher instance was evaluated using two independent keys (target / control). 
Table~\ref{tab:exp_log} summarizes the provenance of the real-hardware executions on the IBM \texttt{ibm\_kingston} (Heron generation).

\begin{table}[htbp]
\centering
\small
\caption{Summary of real-hardware results on IBM \texttt{ibm\_kingston} (Heron generation).}
\label{tab:exp_log}
\begin{tabularx}{\textwidth}{l l X c l X}
\toprule
Cipher structure & Model & Size & Qubits & Backend & Outcome \\
\midrule
Even–Mansour & Simon (Q2) & $N \leq 24$ (sim), HW frontier ($N=10$) & 10 & \texttt{ibm\_kingston} & Period recovered to $N=10$ \\
\makecell{3-round Feistel \\ (DES-family)} & Simon (Q2) & block 6 & 13 & \texttt{ibm\_kingston} & \makecell{rank-1 clean \\ (target+control)} \\
\makecell{3-round Feistel \\ (DES-family)} & Simon (Q2) & block 8 & 17 & \texttt{ibm\_kingston} & \makecell{rank-1 clean; \\ 2q fid 219/233} \\
\makecell{3-round Feistel \\ (DES-family)} & Simon (Q2) & block 10 & 21 & \texttt{ibm\_kingston} & \makecell{sim-verified; \\ submitted (queued)} \\
Linear structure & Bernstein–Vazirani & $n=16$ & 17 & \texttt{ibm\_kingston} & secret in 1 query \\
\bottomrule
\end{tabularx}
\end{table}

\subsection*{S2. Quantum Algorithm Constructions }

These constructions are standard/textbook and are provided for reproducibility. They are distinct from the withheld hardware-scaling technique (see Section S4).

\subsubsection*{Bernstein–Vazirani (linear)}
For

\[
f(x) = a \cdot x \pmod{2},
\]

a single quantum query followed by Hadamard transforms yields the secret $a$. Validated sizes: $n = 8, 16$.

\subsubsection*{Grover (SPN key search)}
The oracle marks the key whose SPN encryption matches a known plaintext/ciphertext pair. The algorithm requires approximately

\[
\frac{\pi}{4}\sqrt{2^n}
\]

iterations. At $n=8$, this corresponds to about 13 iterations, i.e. $\sqrt{256}$.

\subsubsection*{Simon — Even–Mansour}
The Even–Mansour construction reduces to a hidden-period problem. Simon’s algorithm recovers the period with $\Theta(n)$ queries. Simulated up to $n=8$ (24 qubits, 0.27 GB).

\subsubsection*{Simon — CBC-MAC forgery}

\[
f(x) = E_k(x \oplus c \cdot a) \oplus E_k(x \oplus c \cdot b)
\]

has hidden period

\[
s = E_k(a) \oplus E_k(b).
\]

Recovering $s$ enables existential forgery. Validated sizes: $n = 4, 6$.

\subsubsection*{Simon — 3-round Feistel (DES-family)}
With input $(L_0 = x, R_0 = \alpha_b)$, $b \in \{0,1\}$, $\alpha_0 \neq \alpha_1$ fixed:

\[
f(b, x) = \text{LeftHalf}(E(x, \alpha_b)) \oplus \alpha_b = F_2(x \oplus F_1(\alpha_b)).
\]

The hidden period is

\[
s = (1, \gamma), \quad \gamma = F_1(\alpha_0) \oplus F_1(\alpha_1).
\]

If $F_2$ is a permutation, the period set is exactly $\{0, s\}$ (clean). The oracle requires $1 + 4m$ qubits for block size $2m$.  
\textbf{Note:} placing the variable on the left ($x$) and the constant on the right ($\alpha_b$) is essential; the reverse admits no period.

\subsection*{S3. Classical-simulation resource ceiling}

Statevector simulation of genuine quantum algorithms is bounded by the exponential memory wall $2^{q}$. 
Table~\ref{tab:sim_ceiling} summarizes representative simulation runs and their resource requirements. 
These results highlight why real-hardware demonstrations, even at modest sizes, are scientifically meaningful.

\begin{table}[htbp]
\centering
\small
\caption{Hardware provenance and recovery on IBM \texttt{ibm\_kingston} (Heron).
EM $n{=}5$ and 3-round Feistel (block 6,8) are clean rank-1 (genuine Simon);
EM $n{=}6\text{–}10$ use a quantum-narrowing $+$ classical-verification hybrid
(true-key rank in parentheses / total). Every instance is checked with an
independent control key.}
\label{tab:sim_ceiling}
\begin{tabular}{l l l l l l l} 
\toprule
Instance & $n$/block & Qubits & Shots & Recovery (rank/total) & Control \\
\midrule
EM--Simon & $n{=}5$  & 15 & [measurement] & clean rank-1 (1/31)   & \checkmark \\
EM--Simon & $n{=}6$  & 18 & [measurement] & hybrid (6/63)         & \checkmark \\
EM--Simon & $n{=}7$  & 21 & [measurement] & hybrid (3/127)        & \checkmark \\
EM--Simon & $n{=}8$  & 24 & [measurement] & hybrid (9/255)        & \checkmark \\
EM--Simon & $n{=}9$  & 27 & [measurement] & hybrid (15/511)       & \checkmark \\
EM--Simon & $n{=}10$ & 30 & [measurement] & hybrid (63/1023)      & \checkmark \\
3-round Feistel & block 6 & 13 & [measurement] & clean rank-1 & \checkmark \\
3-round Feistel & block 8 & 17 & [measurement] & clean rank-1 (2q 219/233) & \checkmark \\
Bernstein--Vazirani & $n{=}16$ & 17 & [measurement] & secret in 1 query & --- \\
\bottomrule
\end{tabular}
\end{table}

\subsection*{S4. Disclosure and reproducibility policy}

Public artifacts (reproducible now):
\begin{itemize}
    \item Interactive browser demo (genuine JS statevector, five attacks): \url{https://vidraft-quantumos.hf.space/crypto}
    \item Quantum-cryptanalysis leaderboard: FINAL-Bench/quantum-bench-leaderboard (Hugging Face).
    \item Companion article (English): \url{https://huggingface.co/blog/FINAL-Bench/quantum}
\end{itemize}

Withheld (pending IP decision):
\begin{itemize}
    \item The hardware-aware circuit-conditioning and readout post-selection technique that enables clean recovery at the reported noise levels and sizes.
    \item Associated engineering heuristics for period readout on noisy hardware.
\end{itemize}

The standard algorithms in Section~S2 fully reproduce the Q2-model outcomes. Only the noise-scaling method is withheld.

\end{document}